\documentclass[accepted]{uai2023} % after acceptance, for a revised
\usepackage[american]{babel}
\usepackage{natbib} % has a nice set of citation styles and commands
\usepackage{mathtools} % amsmath with fixes and additions
\usepackage{booktabs} % commands to create good-looking tables
\usepackage{tikz} % nice language for creating drawings and diagrams
\usepackage{pifont}
\usepackage{bm}

\title{Causal Discovery for fMRI data: Challenges, Solutions, and a Case Study}

\author[1]{\href{mailto:<rawls017@umn.edu>?Subject=Your UAI 2023 paper}{Eric~Rawls}{}}
\author[1]{Bryan~Andrews}
\author[1,2]{Kelvin~Lim} %is this the correct way to include two affiliations?? Yes
\author[3]{Erich~Kummerfeld}

\affil[1]{%
    Psychiatry \& Behavioral Sciences\\ %Or whatever the full name is
    University of Minnesota\\
    Minneapolis, Minnesota, USA
}
\affil[2]{% Do we have additional affiliations? The VA maybe? Yes, Kelvin asked to include the VA
    Minneapolis VA Health Care System\\
    Geriatrics Research Education and Clinical Center (GRECC)\\
    Minneapolis, Minnesota, USA
  }
\affil[3]{%
    Institute for Health Informatics\\
    University of Minnesota\\
    Minneapolis, Minnesota, USA
}

\begin{document}
\maketitle
\begin{abstract}
Designing studies that apply causal discovery requires navigating many researcher degrees of freedom. This complexity is exacerbated when the study involves fMRI data. In this paper we (i) describe nine challenges that occur when applying causal discovery to fMRI data, (ii) discuss the space of decisions that need to be made, (iii) review how a recent case study made those decisions, (iv) and identify existing gaps that could potentially be solved by the development of new methods. Overall, causal discovery is a promising approach for analyzing fMRI data, and multiple successful applications have indicated that it is superior to traditional fMRI functional connectivity methods, but current causal discovery methods for fMRI leave room for improvement.
\end{abstract}

%our TL;DR: Current causal discovery methods have room for improvement when it comes to fMRI data, however previous applications indicate that despite many challenges they are superior to traditional fMRI connectivity methods.

\section{Introduction: fMRI Brain Data and Effective Connectivity}\label{sec:intro}

Functional Magnetic Resonance Imaging (fMRI) offers the highest-currently-available spatial resolution for three-dimensional images of real-time functional activity across the entire human brain \citep{Glasser2016-gh}. For this reason, enormous resources have been spent to collect fMRI data from hundreds of thousands of individuals for research purposes \citep{Volkow2018-jv,Elam2021-ob,Alfaro-Almagro2018-mp}. This data is collected and analyzed with the purpose of answering a large variety of scientific questions, such as understanding drivers of adolescent substance use initiation \citep{Volkow2018-jv}.

In this paper we focus on questions related to how activity in different areas of the brain may causally influence activity in other areas of the brain \citep{Friston2009-ce}. This can serve a variety of purposes, such as guiding medical interventions like non-invasive brain stimulation (NIBS) \citep{Horn2020-wj}. For example, deep-brain stimulation targets the subthalamic nucleus (STN), however current NIBS technologies cannot directly manipulate activity in STN \citep{Horn2017-gk}. The area could potentially be indirectly manipulated through other brain areas, however this requires learning, either for the population or for each individual, which other brain areas have the greatest causal influence on STN. Standard fMRI analysis unfortunately eschews estimating causal effects.

Current standard practice in fMRI analysis is to describe the connections between brain areas with empirical correlations \citep{Biswal1995-yw}. Partial correlation methods such as \emph{glasso} \citep{Friedman2008-yk} are also used but are less common \citep{Marrelec2006-av}. The connections learned by such methods are called ``functional connections”. This term is used to distinguish them from ``effective connections” where one area is described as causally influencing another \citep{Friston2009-ce,Reid2019-kd,Pearl2000-gd}. Despite the overtly non-causal nature of functional connectivity, it is still used by the larger fMRI research community to identify brain areas that should be targeted with interventions, and more generally to describe how the brain functions. A small but growing community of fMRI researchers are adopting causal discovery analysis (CDA) instead \citep{Spirtes2000-od,Camchong2023-iv,Sanchez-Romero2021-hb,Rawls2022-qv}. These papers have demonstrated that CDA is capable of recovering information from fMRI data above and beyond what is possible with traditional fMRI connectivity methods, however the details of their approaches vary substantially.

In any project involving CDA, the researcher faces many choices (degrees of freedom). Using CDA requires navigating a wide range of algorithms, properties, assumptions, and settings. Previous fMRI CDA studies have often differed in their specifics, but share a number of common strategies. We focus this paper’s discussion on the decisions made during one project: the development and application of the Greedy Adjacencies with Non-Gaussian Orientations (GANGO) method to data from the Human Connectome Project (HCP) \citep{Rawls2022-qv}.

The aim of \cite{Rawls2022-qv} was to describe the structure of effective connectivity in the brain commonly found in healthy individuals while at rest, a.k.a. the resting-state causal connectome. This model could then be used to identify causal connectome alterations that may be responsible for psychopathology in people suffering from mental health disorders, as well as to predict the clinical severity of lesions in different brain areas in terms of its impact on the larger causal connectome. To construct this model, however, numerous decisions were made regarding both how to perform the CDA itself and how to clean and process the raw fMRI data. The need for such decisions comes from several challenges that are universal to CDA fMRI studies, which we describe next.

\section{Practical challenges of applying CDA to fMRI data}\label{sec:challenges}

Any research study where CDA is applied to fMRI data will have to confront numerous study design challenges. Here we enumerate 9 challenges that apply to all CDA fMRI studies.

\textbf{C1: Preprocessing.} Raw fMRI data contains numerous artifacts due to a wide variety of physical, biological, and measurement technology factors. For example, there are many artifacts resulting from the fact that the brain is not a rigid, stable object: head motion will obviously impact which locations in space within the scanner correspond to which parts of the brain, but many other less obvious factors such as blinking, swallowing, and changes in blood pressure due to heartbeats, all apply pressure to the brain and causes it to move and change shape. There is a large space of methods for cleaning and preprocessing fMRI data, and these can have a large impact on any fMRI analysis \citep{Parkes2018-fr,Botvinik-Nezer2020-az}. Further, there is an interaction between how the fMRI data is cleaned and preprocessed and what CDA methods are viable. Many popular fMRI preprocessing methods produce Gaussian data, but some popular causal discovery methods require the data to be non-Gaussian \citep{Ramsey2014-dy}. As such, we can not choose the fMRI preprocessing method and CDA method independently of each other: they must be chosen jointly.

\textbf{C2: Cycles.} Brains are known to contain both positive and negative feedback cycles \citep{Sanchez-Romero2019-pb,Garrido2007-nd}. CDA methods that are capable of learning cyclic relationships will thus be preferable, \emph{ceteris paribus}, to CDA methods that can not learn models with cycles. Further, the CDA methods that can accurately learn cycles primarily operate outside the space of Gaussian distributions. However, as already mentioned, many fMRI cleaning methods force the data to be Gaussian, thus making it essentially unusable for those methods.

\textbf{C3: Undersampling.} The sampling rate of fMRI imaging is much slower than the rate at which neurons influence each other. That is, typical image acquisitions only sample the brain about every 1-2 seconds \citep{Daranyi2021-hq} and the Blood Oxygen Level Dependent (BOLD) response does not peak for approximately 5-7 seconds following activation of neurons \citep{Buckner1998-rg}. Meanwhile, pyramidal neurons can fire up to 10 times per second and interneurons may fire as many as 100 times per second \citep{Csicsvari1999-ks}. Some recent CDA research has focused on undersampled time series data \citep{Hyttinen2016-uz,Hyttinen2017-zy,Cook2017-uq,Solovyeva2023-hk}, however the application of these approaches to parcellated fMRI data remains largely unexplored.

\textbf{C4: Latents.} It is plausible that our measured variables are influenced by some unmeasured common causes, such as haptic or interoceptive feedback from the peripheral nervous system, or even inputs from small brain regions that are not included separately in parcellations but typically lumped together, such as the raphe nucleus (serotonin), the locus coeruleus (norepinephrine), and the ventral tegmental area (dopamine).

\begin{figure}
    \centering
    \includegraphics[width=0.8\linewidth]{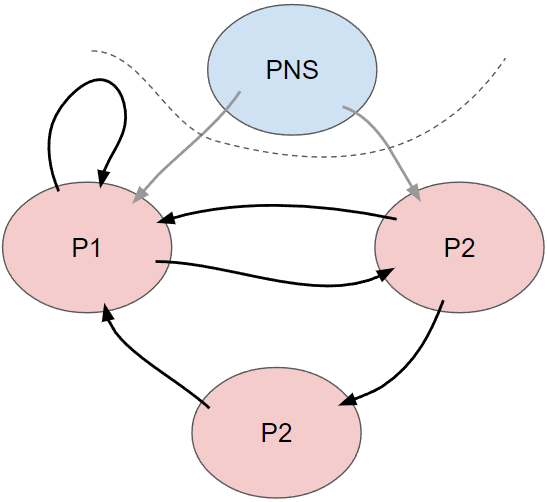}
    \caption{Using CDA on parcellated fMRI data is complicated by the likely presence of both cycles and latent variables. This figure illustrates a toy example of parcels P1, P2, and P3, with unmeasured confounding of P1 and P2 from the peripheral nervous system (PNS). P1 has a cycle of length 1, as activity in P1 directly impacts further activity in P1 (for example, the firing of interneurons in a brain area will suppress the firing of other neurons). There is a 2-cycle between P1 and P2, as each directly causes the other. There is also a 3-cycle between P1, P2, and P3. Many methods capable of learning causal cycles are limited to learning only cycles with length greater than 2, however cycles of all lengths are likely present in brain parcellations.}
    \label{fig:example}
\end{figure}

\textbf{C5: Spatial smoothing.} fMRI is subject to poorly characterized spatial smoothing resulting from the scanner itself and also from standard preprocessing that typically includes spatial smoothing with a Gaussian kernel \citep{Mikl2008-li}. This induces correlations between nearby brain areas. Since these correlations are due only to the measurement and standard preprocessing technologies, they do not reflect causal processes inside the brain. However, current CDA methods will universally attempt to explain these correlations with conjectured causal mechanisms. Performing analysis at the level of parcellations of voxels — biologically interpretable spatially contiguous sets of voxels — is the typical approach that fMRI researchers use to ameliorate this problem.

\textbf{C6: High Dimensionality.} At present, the smallest brain areas that are commonly used for full-brain connectivity analysis are parcellations \citep{Glasser2016-gh,Schaefer2018-wb}. These are smaller than most Regions of Interest (ROIs) or other spatially defined ``brain networks'' (e.g. the ``Default Mode Network''), but much larger than voxels. Individual parcels typically comprise 100-200 voxels or more \citep{Glasser2016-gh}. Voxels are in turn much larger than neurons, containing around 1 million neurons in an 8 mm$^3$ voxel \citep{Ip2021-qk}. There are multiple different whole-brain parcellations, but they typically include hundreds of parcels. Some examples include the recent multimodal parcellation of the human cortex into 360 parcels by \cite{Glasser2016-gh}, and the multiscale parcellation of Schaefer and colleagues that includes between 100 and 1000 parcels \citep{Schaefer2018-wb}. The analysis algorithm must therefore scale to hundreds or several hundreds of variables, which excludes many CDA methods.

\textbf{C7: High Density.} Brain networks are densely connected. On average, the nodes of parcellated fMRI networks are typically connected to at least 10 other nodes \citep{Rawls2022-qv}. With some recent exceptions, most CDA methods have reduced performance and greatly increased computational cost on models with such high density \citep{Lam2022-kt}.

\textbf{C8: Scale-free structure.} Brain networks are scale-free at many different resolutions, including at the resolution of fMRI parcellations \citep{Watts1998-vw,Rawls2022-qv}. This small-world type of connectivity is characterized by having a small number of extremely well-connected nodes, aka hubs. These hubs play critical roles in organizing complex brain functions where multiple regions interact \citep{Van_den_Heuvel2013-di,Crossley2014-yx}. For example, the anterior cingulate cortex (ACC) is densely interconnected with other subcortical and cortical regions, receiving information about emotion and valuation from subcortical brain systems and sending information about the need for control to other brain regions. The extremely high connectivity of these nodes may make them more difficult for some CDA methods to learn, especially as many methods encode sparsity biases that prefer models with more distributed connectivity (like Erdos-Renyi models) \citep{Erdos1960-ex,Karonski1997-qo}.

\textbf{C9: Limited Samples.} fMRI brain data from a single session typically has sample sizes (number of images/frames) that range from several hundred to two thousand \citep{Volkow2018-jv,Elam2021-ob,Alfaro-Almagro2018-mp}. Modern imaging protocols involve imaging at a rate of about 1 capture per second, and the participant is required to stay extremely still. This includes not swallowing and not blinking too much. This is not a comfortable experience, making the total duration of scanning, and thus the total sample size from a single session, necessarily limited. Methods that require more than a few thousand data points are thus not feasible for fMRI, unless they are intended for analyzing multiple sessions or subjects (and thus not focused on our goal of modeling an individual person’s causal connectome at a point in time).

To summarize, the ideal analysis will (1) preprocess the data in a way that cleans as many artifacts as possible while enabling the chosen CDA method, (2) be able to recover causal cycles, (3) be relatively unaffected by or explicitly model temporal undersampling, (4) allow for the possibility of unmeasured confounding, (5) not become biased by spatial smoothing, (6-7) scale to, and retain strong performance on, data with hundreds of densely-connected (average degree 10 or more) variables, (8) retain strong performance for hub nodes in scale free models, (9) achieve all of the above on data with between hundreds and two thousand samples.

\section{CDA and effective connectivity methods for fMRI}\label{sec:CDAmethods}

When selecting a CDA method for fMRI analysis, the large number of algorithms developed over the last thirty years can be intimidating. However, the majority of CDA methods do not meet the minimum requirements for analyzing parcellated fMRI data. For example, many CDA methods do not scale to problems with hundreds of variables (C6). Moreover, methods that rely on the cross-temporal relationships in the data, such as Granger causality \citep{granger1969investigating,friston2014granger}, will not produce meaningful results due to undersampling \citep{barnett2017detectability} (C3). In practice, researchers usually treat fMRI data as if it were independent and identically distributed while applying CDA methods to avoid this issue. Below, we review popular CDA methods that are appropriate for analyzing parcellated fMRI data. Table \ref{tab:CDA_algos} compares these methods relative to the 9 challenges.

\begin{table*}[ht]
    \centering
    \begin{tabular}{rccccccccc}
         & C1 & C2 & C3 & C4 & C5 & C6 & C7 & C8 & C9 \\
        GES & \ding{51} & \ding{55} & \ding{55} & \ding{55} & \ding{55} & \ding{51} & \ding{55} & \ding{55} & \ding{51} \\
        GRaSP & \ding{51} & \ding{55} & \ding{55} & \ding{55} & \ding{55} & $\bm \sim$ & \ding{51} & \ding{51} & \ding{51} \\
        Direct LiNGAM & $\bm \sim$ & \ding{55} & \ding{55} & \ding{55} & \ding{55} & \ding{51} & \ding{51} & \ding{51} & \ding{51} \\
        GANGO & $\bm \sim$ & $\bm \sim$ & \ding{55} & \ding{55} & \ding{55} & \ding{51} & $\bm \sim$ & $\bm \sim$ & \ding{51} \\
        FASk & $\bm \sim$ & \ding{51} & \ding{55} & \ding{55} & \ding{55} & \ding{51} & $\bm \sim$ & $\bm \sim$ & \ding{51} \\
        Two-Step & $\bm \sim$ & \ding{51} & \ding{55} & $\bm \sim$ & \ding{55} & \ding{51} & $\bm \sim$ & $\bm \sim$ & \ding{51} \\
    \end{tabular}
    \caption{A comparison of the discussed CDA methods relative to the 9 challenges where (\ding{51}) denotes the challenge is addressed, $(\bm \sim)$ denotes the challenge is partially addressed, and (\,\ding{55}\,) denotes the challenge is not addressed. Note that while the same challenge may be addressed by multiple algorithms, it is still possible that some algorithms address have better performance on this challenge than others. For example, GES, GRaSP, Direct LiNGAM, and GANGO appear to have better accuracy on low samples (C9) than FASk and Two-Step.}
    \label{tab:CDA_algos}
\end{table*}

\textbf{Greedy Equivalent Search (GES)} is a two-phase greedy search algorithm that moves between equivalence classes of Directed Acyclic Graphs (DAGs) by adding or removing conditional independence relations. Fast Greedy Equivalent Search (fGES) is an efficient implementation of GES capable of scaling up to a million variables (C6). It does not have any special preprocessing requirements (C1), and has good model performance on limited samples (C9). However, this algorithm is intended for sparsely connected networks and suffers in terms of performance and scalability on densely connected problems (C7) \citep{chickering2002optimal,Ramsey2017-tg}.

\textbf{Greedy Relaxations of the Sparsest Permutation (GRaSP)} is a hierarchy of greedy search algorithms that move between variable orderings. For brevity, we will use the GRaSP acronym to refer to the most general algorithm in the hierarchy. Starting from a random order, GRaSP repeatedly iterates over pairs of variables that are adjacent in the DAG constructed by applying the Grow Shrink (GS) \citep{margaritis1999bayesian} in a manner consistent with the order, modifying the order in a way consistent with flipping the corresponding edge in the DAG. GRaSP does not have any special preprocessing requirements (C1), and has good model performance on limited samples (C9). However, unlike fGES, GRaSP retains its good performance on high density models (C7) but only scales to one or two hundred variables (C6) \citep{lam2022greedy}.

\textbf{Direct Linear Non-Gaussian Acyclic Model (LiNGAM)} is a greedy algorithm that constructs a variable ordering based on a pairwise orientation criterion \citep{Hyvarinen2013-bj}. Starting from an empty list, the order is constructed by adding one variable at a time, the one that maximizes the pairwise orientation criterion, until a full ordering is constructed. Once the order is constructed, it is projected to a DAG. The pairwise orientation criterion uses measures non-Gaussianity so some preprocessing techniques cannot be used (C1). Moreover, LiNGAM cannot learn cycles (C2). That being said, the method scales fairly well and has not problem learning scale-free structures (C6), (C7), and (C9) \citep{shimizu2011directlingam}.

The next three methods broadly fall into the same class of algorithms. These approaches are two stepped approaches where the first step learns a graph using an existing CDA method and the second step augments and (re)orients that edges of the graph learned in the first step; this general approach was pioneered by \cite{hoyer2008causal}. All three of these methods use non-Gaussianity and thus some preprocessing techniques cannot be used (C1). Moreover, the properties of these methods are impacted their algorithm choice in the first step, for example, if the chosen fails to scale in some aspect, then so will the overall method (C6 - C9).

\textbf{Greedy Adjacencies with Non-Gaussian Orientations (GANGO)} uses fGES \citep{Ramsey2017-tg} as a first step in order to learn the adjacencies and then uses the RSkew pairwise orientation rule \citep{Hyvarinen2013-bj}, also referred to as robust skew, for orientations. This method allows for cycles (with the exception of two-cycles) and scales well \citep{Rawls2022-qv}.

\textbf{Fast Adjacency Skewness (FASk)} uses fast adjacency search (FAS), which is the adjacency phase of the PC algorithm, as a first step in order to learn the adjacencies and then uses a series of tests to add additional adjacencies, orient two-cycles, and orient directed edges. This method allows for cycles and scales well \citep{Sanchez-Romero2019-pb}.

\textbf{Two-Step} uses adaptive lasso or FAS as a first step in order to learn the adjacencies and then uses independent subspace analysis (ISA) or independent component analysis (ICA) if no latent confounders are identified. This method allows for cycles and latent confounding, and scales well \citep{Sanchez-Romero2019-pb}.

%\textbf{Regression Dynamic Causal Modeling (rDCM)} is a dynamical system motivated by biology and simplified for efficiency, fit using Variational Bayes (VB) \citep{Frassle2017-sg, Frassle2018-fu, Frassle2021-cn}. While its properties as a causal discovery algorithm remain relatively unstudied, rDCM is another option for researchers pursuing effective connectivity models from fMRI data.

\section{Interdependencies among CDA and fMRI processing methods}\label{sec:interdep}

This section briefly covers some of the more important ways in which choice of CDA method and choice of fMRI processing methods interact, and how these complexities can be successfully navigated.

First, as discussed above, the better CDA methods for learning cycles reliably requires using non-Gaussian statistics, however many preprocessing methods force the fMRI data to conform to a Gaussian distribution. While the Cyclic Causal Discovery (CCD) algorithm \citep{Richardson1996-us} can recover causal graphs with cycles from Gaussian data, this algorithm performs poorly on finite samples and is rarely used. Methods exploiting non-Gaussian structure in BOLD data achieve higher precision and recall with simulated BOLD data [19]. Fortunately, there are approaches to preprocessing fMRI that do not completely remove the non-Gaussian signal. Those approaches are thus recommended for using CDA on fMRI data.

Preprocessing removes artifacts and recovers physiological brain signals via some combination of temporal filtering, spatial smoothing, independent components analysis (ICA), and confound regression \citep{Glasser2013-zm}. Some of these steps, particularly temporal filtering, can drastically modify the data distribution. For example, \citep{Ramsey2014-dy} demonstrated that certain high-pass temporal filters made parcellated fMRI time series more Gaussian. This effect was particularly strong for Butterworth filters, which were applied in \citep{Smith2011-da}. As such, it is likely that the results of \citep{Smith2011-da} were unrealistically pessimistic with regards to methods that assume non-Gaussianity. This effect holds true for the filter built into the Statistical Parametric Mapping (SPM) software, while being negligible for the filter built into the fMRIB Software Library (FSL) software. Thus, high-pass filtering including the specific software and filter used is a critical point of attention during data preparation for CDA.

For filtered data that maintain non-Gaussianity in the BOLD signal, it's crucial to confirm the data meet distributional assumptions. A recent cortex-wide human causal connectome analysis discovered that minimally preprocessed cortical BOLD signal was non-Gaussian for all subjects \citep{Rawls2022-qv}. In that same dataset, however, the subcortical parcel time series are not non-Gaussian. This could stem from Gaussian noise corrupting non-Gaussian BOLD activity, especially since subcortical regions typically display low signal-to-noise ratios. There is potential for enhancing BOLD data's compatibility with CDA methods by employing newer techniques that eliminate Gaussian noise, like the NORDIC method \citep{Moeller2021-wd,Vizioli2021-ta}, which suppresses Gaussian thermal noise from high-resolution scan parameters. However, to date we are unaware of any studies that pursue this combination of preprocessing methods and CDA.

\section{Additional complications of CDA on fMRI}\label{sec:complications}

Regarding the challenges of high dimensionality and limited samples, this area fortunately has numerous causal discovery solutions. All of the methods discussed in Section \ref{sec:CDAmethods} are capable of scaling to the number of parcels found in the most widely used parcellations, while maintaining good performance (although both runtime and performance can vary substantially across these methods).

The structural challenges of high-density and scale-free models also have some solutions. In particular, recently developed permutation-based methods such as GRaSP and BOSS both retain their high performance as model density increases. These methods have increased computational cost compared to faster methods like fGES, but can still scale comfortably to hundreds of parcels, even on a personal computer. Most other methods appear to have substantial drops in performance as density increases, so using one of the few methods that tolerates high density models is recommended.

Since fMRI preprocessing may leave non-Gaussian marginal distributions, it’s worth considering whether the CDA methods that assume a linear-Gaussian model still perform well. In general, such methods retain their performance for edge adjacencies \citep{Smith2011-da}, while their performance on edge orientations has mixed results. There are known cases where the orientations become essentially random \citep{Smith2011-da}, while we have observed other cases where the orientations only exhibit a slight drop in accuracy.

Collectively, the challenges and available tools point towards a particular approach: 
\begin{enumerate}
    \item to ensure sample size is not too small (C9), analyze data from study protocols that allow for adequate scanner time for each individual session;
    \item for preprocessing (C1), remove as many fMRI artifacts as possible while retaining as much non-Gaussianity in the marginals of the parcellated time-series data as possible; 
    \item for undersampling (C3) and spatial smoothing (C5), use a cross-sectional approach to analyze parcelations; 
    \item due to high dimensionality (C6), high density (C7), and scale-free brain structure (C8), use a scalable high-density-tolerant method to learn the skeleton (adjacencies) of the parcels;
    \item in order to learn cycles (C2) use a non-Gaussian orientation method to re-orient edges; 
    \item In consideration of possible latent confounding (C4), one can either 
    \begin{enumerate}
        \item use a scalable CDA method capable of learning both cycles and latent confounding, or 
        \item focus the primary results on aggregate features of the model, such as connections among multiple parcelations in shared networks, rather than on individual edges.
    \end{enumerate}
\end{enumerate}

The next section reviews a project where this general approach was taken.

\section{Case Study: Application of this approach to the Human Connectome Project (HCP)}
\label{sec:casestudy}

A previous project can serve as an example of the above thought process \citep{Rawls2022-qv}. In that project, the authors made the following considerations with respect to the 9 challenges. 

\begin{figure*}
    \centering
    \includegraphics[width=0.95\textwidth]{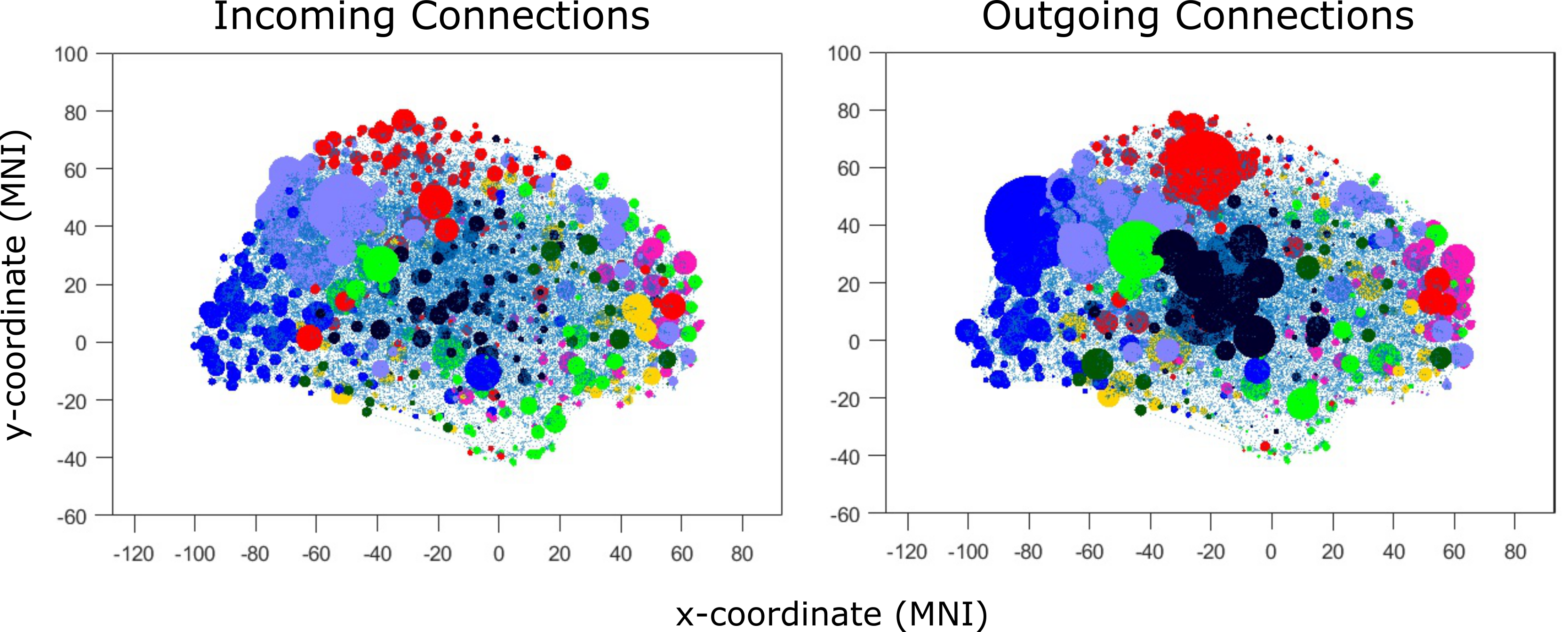}
    \caption{Application of CDA methods to BOLD data allows estimating patterns of directed connectivity, separating incoming and outgoing connections. In these brain images, circles are plotted at the 3D centroids of cortical parcels and the size of the circle indicates the number of connections (the degree) of the parcel. Individual causal connectomes reveal a striking dissociation between brain regions with especially high indegree and outdegree. Different colors refer to different brain functional communities (resting-state networks).}
    \label{fig:inoutdegree}
\end{figure*}

\textbf{Challenge 1 (preprocessing):} The minimal HCP processing pipeline \citep{Glasser2013-zm} was used, to conserve as much non-Gaussian signal as possible and enable the use of non-Gaussian CDA methods for learning cycles. Non-Gaussianity of data were statistically verified by simulating surrogate Gaussian data for comparison.

\textbf{Challenge 2 (cycles):} RSkew was used to re-orient edges \citep{Hyvarinen2013-bj} after confirming non-Gaussianity of the preprocessed cortical parcellations, enabling discovery of cycles involving three or more variables.

\textbf{Challenge 3 (undersampling):} The time series element of the data was ignored, and the parcellated time series were instead analyzed as cross-sectional data. While this approach does not make use of the available time-order information, it avoids relying on the heavily undersampled time dimension of the data.

\textbf{Challenge 4 (latents):} No effort was made to directly model or account for latent variables. Findings were reported at an aggregate level rather than individual edges.

\textbf{Challenge 5 (spatial smoothing):} These data were only minimally smoothed (2 mm in surface space) \citep{Glasser2013-zm}, thus excessive smoothing was not introduced in the data. In addition, the parcellated time series was analyzed rather than voxels, which further reduced the impact of smoothing.

\textbf{Challenge 6 (high dimensionality):} A 360-node cortical parcellation was used \citep{Glasser2016-gh}. FGES was used for the more computationally intensive adjacency search, which is among the most scalable methods for CDA \citep{Ramsey2017-tg}. 

\textbf{Challenge 7 (high density):} The study used fGES, which scales well but can have lower performance for extremely dense graphs. Better performance for dense brain graphs could potentially be achieved by applying a high-density-tolerant algorithm such as GRaSP \citep{Lam2022-kt}.

\textbf{Challenge 8 (scale free):} \cite{Rawls2022-qv} reported the existence of nodes that were more highly connected than expected under chance, which is characteristic of scale-free networks. See Figure \ref{fig:scalefreesmallworld}. However, for especially highly-connected hub regions, some methods such as GRaSP might provide higher precision for assessing scale-free structure in future studies.

\textbf{Challenge 9 (limited samples):} The HCP collected two runs of fMRI per day of 1200 images per run. The study applied CDA to concatenated standardized time series from these two runs, thus the number of samples was extremely high (2400 total). The CDA methods that were selected are also known to have good finite-sample performance.

\textbf{Overall:} This recent large-scale application of CDA for deriving individualized causal connectomes addressed many of the challenges we identified. However, the challenges of high density and scale-free connectivity could potentially be better addressed by applying newer permutation-based CDA methods \citep{lam2022greedy}. Several challenges, such as limited samples, spatial smoothing, and preprocessing, were partially or entirely solved by the specific data set the method was applied to, and might pose problems in other data sets.

\subsection{Results from the HCP Case Study}

\begin{figure*}
    \centering
    \includegraphics[width=0.65\textwidth]{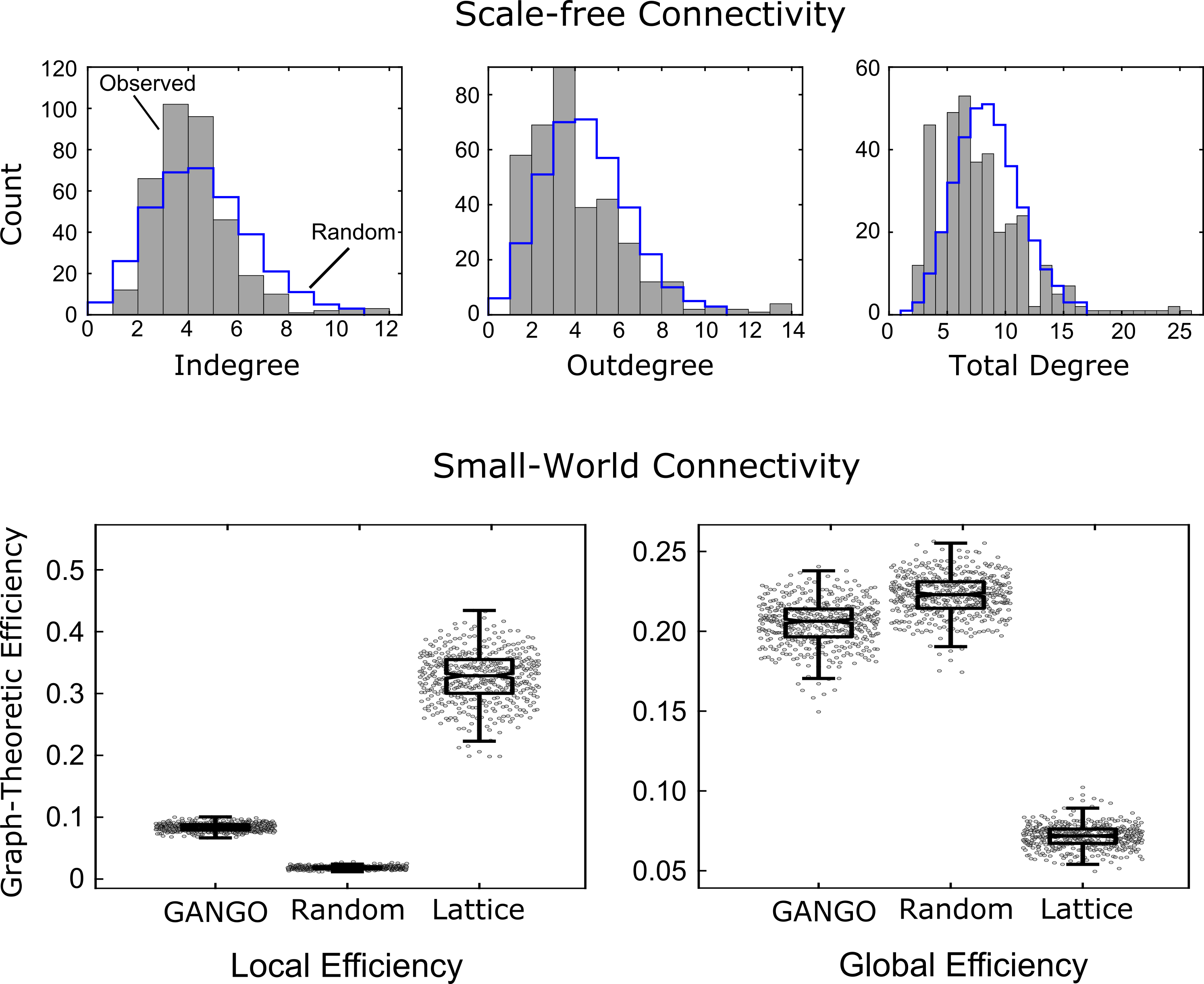}
    \caption{The case study that developed the GANGO framework found that the causal connectomes produced were both scale-free and small-world graphs. The scale-free nature of GANGO connectomes is demonstrated by degree distributions that were more skewed than random graphs. The small-world nature of GANGO connectomes is demonstrated by global efficiency that is nearly equal to random graphs but higher than lattice graphs, combined with local efficiency that is greater than random graphs. Panels illustrating scale-free connectivity are adapted with permission from \cite{Rawls2022-qv}.}
    \label{fig:scalefreesmallworld}
\end{figure*}

Here we briefly review the findings from the HCP case study by \cite{Rawls2022-qv}. This study developed the GANGO causal connectivity method, which was applied to n=442 resting-state fMRI data sets. The connectomes produced were extremely sparse (2.25\% edge density) compared to Pearson correlation connectomes, which are often thresholded to an edge density of 5-50\%. Nevertheless, graphs produced by GANGO were fully connected in nearly all cases, which was not the case for standard Pearson correlation graphs.

The GANGO method produced graphs with a scale-free degree distribution. More specifically, the degree distributions were skewed by existence of hub nodes with very high connectivity (some with total degree exceeding 20). See Figure \ref{fig:scalefreesmallworld}. These hub nodes were disproportionately concentrated in brain networks tied to attention and executive control, while Pearson correlations instead emphasized hub connectivity of early sensory regions. Graphs produced by the GANGO method also show small-world connectivity, with global efficiency nearly as high as random graphs but local efficiency much higher than random graphs. Overall, this case study showed that a causal discovery algorithm specifically designed to meet the unique challenges of fMRI data recovers physiologically plausible connectomes with small-world and scale-free connectivity patterns characteristic of biological networks.

\section{Research gaps and promising future directions}

We have outlined nine challenges researchers will face when attempting to apply CDA to parcellations of fMRI data, as well as some available CDA technologies and their ability to overcome those challenges. The case study discussed in Section \ref{sec:casestudy} attempted to use a mixture of strategies to overcome those challenges, but many challenges remained only partially addressed, or were even largely ignored. In this section, we review the current research gaps as we perceive them, and point towards future directions to empower future applications of CDA to better elucidate the brain’s causal connectome for both scientific and medical purposes.

\textbf{Gap 1: CDA methods for high-dimensional, high-density, scale-free models.} Previous work \citep{Lam2022-kt} has shown that many popular CDA methods unfortunately do not perform well when nodes have larger numbers of connections to other nodes. The only currently published method that appears to overcome this limitation is limited to about 100 variables, which is substantially fewer than most parcellations \citep{Glasser2016-gh,Schaefer2018-wb}. We are aware of research on new algorithms and CDA implementation technologies that may overcome this gap, however that work has not yet been published. However, such methods could be used as a replacement for methods like fGES in future fMRI studies.

\textbf{Gap 2: Reliance on skewed data.} While it is the case for some fMRI data that minimal preprocessing is able to retain statistically significant skew, we are also aware of other fMRI data where even after only minimal preprocessing the data are not significantly skewed. One possible future direction would be to incorporate additional information from higher moments such as kurtosis to ensure that the non-Gaussian orientation methods can be used as much as possible. Another approach would be to distill as much non-Gaussian signal as possible using a method like independent components analysis (ICA) \citep{comon1994independent} to construct a new set of features from the parcellated time series, and then perform CDA on the maximally non-Gaussian components of each parcel instead.

\textbf{Gap 3: Latent variables.} While there exist CDA methods that do not assume causal sufficiency, and thus can tolerate and even identify unmeasured common causes, they generally have significant difficulties with other challenges. For example, the standard methods for handling latent variables, like FCI and GFCI, do not allow for cycles, have limited scalability, and perform poorly for high-density models. The Two-Step algorithm \citep{Sanchez-Romero2019-pb} can in theory incorporate unmeasured confounding in its models, however we are not aware of any theoretical or practical evaluation of its performance in the presence of unmeasured confounding.

\textbf{Gap 4: Extension to other brain imaging technologies.} Future exploration should expand CDA methodology to neural temporal data beyond BOLD signals (fMRI data). For example, electroencephalography (EEG) provides a dynamic view of brain activation with exceptional temporal resolution. Current EEG causal connectivity analysis techniques, such as Granger causality, are fruitful yet limited. These current methods do not differentiate brain oscillations from aperiodic activity, which is critical given recent evidence that aperiodic activity sometimes wholly explains group differences in power spectral density \citep{merkin2023age}. Techniques have recently emerged separating aperiodic and oscillatory contributions to EEG power spectra \citep{donoghue2020parameterizing}, even extending to time-frequency domain for time-resolved separation \citep{wilson2022time}. However, these haven't been incorporated into neural connectivity analyses. EEG data also present challenges in identifying effective connectivity patterns due to volume conduction -- instant, passive electricity conduction through the brain separate from actual neural interactions, resulting in non-independent EEG sensor-level estimates \citep{Nunez2006}. This non-independence hampers CDA, necessitating volume conduction removal from data before connectivity estimation (EEG preprocessing). We suggest these obstacles could be mitigated by first removing volume conduction from EEG data, then separating aperiodic and oscillatory spectral contributions. Applying CDA to isolated EEG oscillatory power estimates could potentially reveal effective connectivity patterns unhindered by aperiodic activity.

\section{Conclusion}

Here we have outlined nine challenges that will be faced by researchers attempting to use CDA for fMRI effective connectivity analysis, and presented a recent case study that attempted to resolve at least some of these challenges. We have also discussed challenges that remain following this case study, such as the continuing search for CDA methods that can discover densely connected graphs and hub nodes with especially high connectivity resulting from scale-free connectivity profiles.

In summary, there are a number of decisions faced by researchers who hope to use CDA for fMRI analysis. By openly discussing these researcher degrees-of-freedom and a recent attempt to resolve these decisions, we hope to foster continued interest in and engagement with the idea that CDA can provide a data-driven method for reconstructing human causal connectomes.

\begin{contributions} % will be removed in pdf for initial submission 
					  % (without ‘accepted’ option in \documentclass)
                      % so you can already fill it to test with the
                      % ‘accepted’ class option
    %Briefly list author contributions. 
    %This is a nice way of making clear who did what and to give proper credit.
    %This section is optional.

    %H.~Q.~Bovik conceived the idea and wrote the paper.
    %Coauthor One created the code.
    %Coauthor Two created the figures.
    All authors contributed to the ideas in this paper and contributed to its drafting. All authors have reviewed and edited the paper's content for correctness.
\end{contributions}

\begin{acknowledgements} % will be removed in pdf for initial submission,
						 % (without ‘accepted’ option in \documentclass)
                         % so you can already fill it to test with the
                         % ‘accepted’ class option
    %Briefly acknowledge people and organizations here.
%
    %\emph{All} acknowledgements go in this section.
    ER was supported by the National Institutes of Health’s National Center for Advancing Translational Sciences, grants TL1R002493 and UL1TR002494. BA was supported by the Comorbidity: Substance Use Disorders and Other Psychiatric Conditions Training Program T32DA037183. EK was supported by grants P50 MH119569 and UL1TR002494.
\end{acknowledgements}

% References
\bibliography{References}
\end{document}